\documentclass[aps,pra,twocolumn,superscriptaddress,10pt,longbibliography]{revtex4-2}

\usepackage{color}
\usepackage{amsmath}
\usepackage{amssymb}
\usepackage{bm}
\usepackage{placeins}
\usepackage{overpic}

\usepackage[T1]{fontenc}

\usepackage{circledsteps}
\usepackage{hyperref}

\usepackage[capitalize]{cleveref}

\usepackage{orcidlink} 
\usepackage{graphicx}
\usepackage{xcolor}

\hypersetup{
  colorlinks=true,
  allcolors = blue,
  pdftitle={}
}

\newcommand{\scref}[2]{\namecref{#1}~\hyperref[#1]{\ref*{#1}#2}}

\begin{document}

\title{Emergent thermal behavior, synchronization, and non-Hermitian phase transitions \\ in driven-dissipative photon condensates}

    \author{Moritz Janning\,\orcidlink{0000-0002-8191-3869}}
    \thanks{These authors contributed equally to this work}
    \email{\\ s6mojann@uni-bonn.de}
    \email{rkramer@uni-bonn.de}
    \affiliation{Physikalisches Institut, Universit\"at Bonn, Nussallee 12, 53115 Bonn, Germany}

    \author{Roman Kramer\,\orcidlink{0009-0002-5018-0188}}
    \thanks{These authors contributed equally to this work}
    \email{\\ s6mojann@uni-bonn.de}
    \email{rkramer@uni-bonn.de}
    \affiliation{Physikalisches Institut, Universit\"at Bonn, Nussallee 12, 53115 Bonn, Germany}

    \author{Aya Abouelela\,\orcidlink{0009-0004-3781-9252}}
    \email{ayaaabou@uni-bonn.de}
    \affiliation{Physikalisches Institut, Universit\"at Bonn, Nussallee 12, 53115 Bonn, Germany}    

    \author{Michael Turaev\,\orcidlink{0009-0003-8566-5078}}
    \email{mturaev@uni-bonn.de}
    \affiliation{Physikalisches Institut, Universit\"at Bonn, Nussallee 12, 53115 Bonn, Germany}
    
    \author{Johann Kroha\,\orcidlink{0000-0002-3340-9166}}
	\email{jkroha@uni-bonn.de}
    \affiliation{Physikalisches Institut, Universit\"at Bonn, Nussallee 12, 53115 Bonn, Germany}
    \affiliation{\hbox{School of Physics and Astronomy, University of St. Andrews, North Haugh, St. Andrews, KY16 9SS, United Kingdom}}
    
    \author{Sayak Ray\,\orcidlink{0000-0003-3944-6715}}
    \email{sayak@uni-bonn.de}
    \affiliation{Physikalisches Institut, Universit\"at Bonn, Nussallee 12, 53115 Bonn, Germany}
    \affiliation{Department of Physics and Astronomy, Aarhus University, Ny Munkegade 120, Aarhus DK-8000, Denmark}

    \begin{abstract}
        Driven-dissipative Bose-Einstein condensates of photons in optical resonators have been shown theoretically to exist as a long-lived, yet metastable state stabilized by a \textit{ghost attractor}, despite continuous driving \href{https://doi.org/10.1103/hcsq-dwcg}{[Phys. Rev. Lett. \textbf{135}, 053402 (2025)]}. Here we analyze the properties of this metastable condensate as well as the relaxation dynamics into and out of this state by means of the Lindblad master-equation approach, treating the condensate and the equal-time, noncondensed fluctuations on the same footing. Despite its nonequilibrium origin, the quasistationary condensate exhibits thermal behavior in that the order-parameter fluctuations scale with system size as in thermodynamic equilibrium, and the photon-number distribution in multimode resonators is in excellent agreement with a Bose-Einstein distribution. A linear stability analysis further reveals the presence of several exceptional points, resulting in multiple non-Hermitian phase transitions associated with the relaxation dynamics into and out of the metastable condensate. These behaviors reconcile apparently contradictory experimental results.
    \end{abstract}
		
\maketitle

\section{Introduction}

Open quantum systems are ubiquitous in nature, since no physical system can be completely isolated from its environment. As a consequence, they experience decoherence and dissipation, and their dynamics violate the unitary time evolution of  Schr\"odinger-like, conserving systems \cite{Petruccione_book}. Driving an open system away from thermodynamic equilibrium gives rise to a rich variety of phenomena, including nonreciprocity, non-Markovian effects, relaxation toward novel stationary states that could not be reached otherwise, and non-Hermitian phase transitions (NHPTs) at so-called exceptional points (EPs) \cite{EP-review, EP-review_2,roccati2026perspective}. An EP is a singularity in configuration space of a non-Hermitian system where two or more eigenmodes coalesce, i.e. their eigenvalues become degenerate and the eigenmodes align with each other \cite{Kato13}. Besides their mathematical significance, EPs have been experimentally realized in a variety of physical platforms, including photonic \cite{EP-review_2} or ultracold gas \cite{Wang24} systems, mesoscopic devices \cite{Wu19} and even in optically pumped, bulk condensed-matter systems \cite{Turaev24}. 

Over the past decade, open photon Bose–Einstein condensates (BECs) realized in dye-filled microcavities \cite{Weitz10} and more recently in semiconductor quantum wells \cite{Nyman24a} and vertical cavity surface emitting laser devices \cite{Pieczarka24}, have emerged as versatile platforms for studying driven–dissipative quantum many-body physics \cite{Ciuti13, Wouter22}. In the grand-canonical or low-density regime, large number fluctuations \cite{Schmitt14} comparable to the average photon number lead to phase jumps \cite{Schmitt16} and a breakdown of temporal coherence \cite{Walker18,Nyman24b}, while in the high-density, canonical regime a coherent photon BEC exists over the entire observation time \cite{Weitz10,Nyman24a,Pieczarka24}. Experiments show an intriguing ambivalence of the driven-dissipative photon BEC between thermal behavior and nonequilibrium features. On one hand, the photon BEC obeys an equilibrium fluctuation-dissipation relation \cite{Schmitt23}. On the other hand, a NHPT was found within the photon BEC phase \cite{Kroha21}, a clear nonequilibrium signature. In that context, recent theoretical work showed that the photon BEC is, in fact, metastable, albeit with a lifetime that typically exceeds the experimental observation time in the canonical regime \cite{Ray25}. It was found that this metastability is due to the presence of a ghost attractor, a fixed point in configuration space which lies outside the physically reachable realm but nevertheless attracts the system's dynamics. As a consequence, the dynamics stall in a plateau-like fashion, and the photon BEC decays exponentially slowly according to a single, small repulsive Lyapunov exponent \cite{Ray25}. The ghost attractor dynamics is well known in nonlinear dynamics \cite{Strogatz-book, Strogatz89, Koch24}, and has recently been identified experimentally in photonic quantum gases \cite{Klars2026}. %It is fundamentally different from many-body prethermalization \cite{Kehrein08, Eckstein11, Essler15, Rigol19, Ray20}.

The purpose of the present article is to clarify the nature of the metastable BEC state and to reconcile the apparent thermal behavior with the nonequilibrium signatures. To that end, we analyze the system in its different dynamical regions. First, we find that during the metastable plateau, the relative order-parameter fluctuations scale with system size $M$ as $1/\sqrt{M}$, with $M$ the number of dye molecules, characteristic for thermal behavior. Second, in multimode resonators we find that the temporal plateau behavior persists as in the single-mode resonators. Furthermore, within the plateau, the complex photon-condensate amplitudes in the different resonator modes perform synchronized oscillations, implying that a single, quantum-coherent condensate exists, while the photon number distribution in the resonator modes is in excellent agreement with a thermal Bose-Einstein distribution. Finally, in addition to the pump-intensity controlled NHPT of the photon density found in \cite{Kroha21}, we find that the photon condensate undergoes a NHPT during the time periods of initial relaxation into as well as relaxing out of the metastable regime. Thus, demonstrating that the thermal and the non-Hermitian (nonequilibrium) behaviors are temporally separated and consistent with each other and with experiments \cite{Weitz10,Kroha21}. 

The paper is organized as follows. After presenting the model and the Markovian rate-equation approach for the open photon gas in a dye-filled optical cavity in \cref{sec:theory}, we remind the reader of the driven dissipative dynamics including their fixed points for experimentally relevant parameters in \cref{sec:dynamics}. \cref{sec:fluctuation} and \cref{sec:multimode} present the analysis of the fluctuations and of the mode distribution, respectively, both showing thermal behavior in the metastable condensate. In \cref{sec:NHPandEP} the relaxation dynamics of the photon condensate into and out of this state are analyzed, which exhibit NHPTs in both cases.  

\begin{figure}[t]
    \centering
    \includegraphics[width=\linewidth]{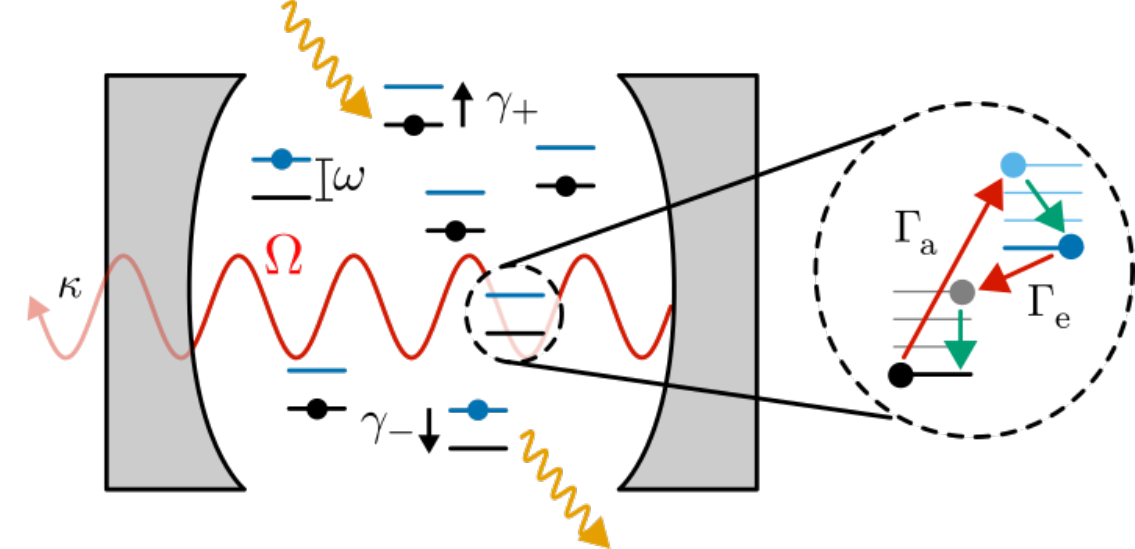}
    \caption{Sketch of the setup. The microcavity with single-mode frequency $\Omega$ is filled with dye molecules of electronic transition frequency $\Delta$. The inset shows fast relaxation of the vibrational modes of molecules (green arrows). The dissipative processes include the phonon-assisted absorption (emission) rates $\Gamma_{\mathrm{a}}$ $(\Gamma_{\mathrm{e}}$), the cavity photon loss $\kappa$, the external pumping and nonradiative decay of the excited molecules $\gamma_{\pm}$, respectively.}
    \label{fig:schematic}
\end{figure}

\section{Model and Dynamical Equations} \label{sec:theory}

Photon Bose-Einstein condensation has been experimentally realized in a system of dye molecules interacting with the photonic modes of a microresonator \cite{Weitz10}. The microscopic dynamics of the system can be described by an extended Tavis-Cummings Hamiltonian \cite{Keeling13, Keeling15, TimBode19}. We restrict ourselves to a single cavity mode which may simultaneously host a photon condensate and noncondensed fluctuations. The extension to multimode resonators is straightforward, and is discussed in \cref{sec:multimode}. Each dye molecule comprises an electronic ground and an excited state that are coupled to vibrational modes (phonons), as illustrated in \cref{fig:schematic} \cite{Keeling13,Keeling15,TimBode19,Ray25}. The phonon subsystem reaches thermal equilibrium at ambient temperature on a time scale much faster than all other scales in the system. This results from the exceedingly large number of vibrational modes relative to  electronic excitations and fast collisions with the solvent, and it can thus be considered as a thermal bath. In addition, the system undergoes incoherent processes due to cavity loss or external incoherent pumping and nonradiative decay of molecule excitations (see \cref{fig:schematic}). The resulting Lindblad master equation for the reduced density matrix $\hat{\rho}$ of photonic and electronic excitations reads \cite{Ray25}, 
\begin{align}
    \partial_t \hat{\rho} &= -i \left[\hat{H}_0,\hat{\rho}\right] + \sum_{j=1}^M \left(\Gamma_{\mathrm{a}} \mathcal{L}[\hat{a} \hat{\sigma}_j^+] + \Gamma_{\mathrm{e}} \mathcal{L}[\hat{a}^\dagger \hat{\sigma}_j^-]\right) \nonumber \\
    &+ \sum_{j=1}^M \left(\gamma_{+} \mathcal{L}[\hat{\sigma}_j^+] + \gamma_{-} \mathcal{L}[\hat{\sigma}_j^-] \right) + \kappa \mathcal{L}[\hat{a}].
\label{eq:LME}
\end{align}
After integrating out the phonon bath by a polaron transformation \cite{Keeling13,Keeling15,TimBode19}, the Hamiltonian $\hat{H}_0$, which governs the coherent dynamics of the system, reads in the rotating frame, 
\begin{equation}
    \hat{H}_0 = \delta \hat{a}^\dagger \hat{a} + g_\beta  \sum_{j=1}^M \left( \hat{a}^\dagger \hat{\sigma}_j^- + \hat{a} \hat{\sigma}_j^+ \right),
    \label{eq:JCH}
\end{equation}
where $\hat{a}$ $(\hat{a}^\dagger)$ is the annihilation (creation) operator of photons in the cavity mode, with $\delta=\Omega-\Delta$ the detuning of the cavity frequency $\Omega$ from the molecule excitation energy $\Delta$ and $g_\beta$ the renormalized Jaynes-Cummings coupling depending on the phonon temperature \cite{Pelster18}. The two-level molecules $j$ are represented by the Pauli matrices $\hat{\sigma}_j^z$, where $j=1,\,2,\,\dots\,M$, together with the raising and lowering operators $\hat{\sigma}_j^\pm = (\hat{\sigma}_j^x \pm i\hat{\sigma}_j^y)/2$. Dissipative processes are described by the Lindblad superoperator $\mathcal{L}[\hat{\mathcal{O}}]$ associated with a jump operator $\hat{\mathcal{O}}$, acting on the density matrix as $\mathcal{L}[\hat{\mathcal{O}}] = \hat{\mathcal{O}} \hat\rho \hat{\mathcal{O}}^\dagger - \left(\hat{\mathcal{O}}^\dagger \hat{\mathcal{O}} \hat\rho + \hat\rho \hat{\mathcal{O}}^\dagger \hat{\mathcal{O}}\right)/2$. In \cref{eq:LME}, $\kappa$ denotes the photon loss rate, while $\gamma_\pm$ represent the incoherent pump and nonradiative decay rates of the molecules, respectively. The phonon-assisted absorption and emission rates of photons by the molecules are given by the Einstein coefficients $\Gamma_{\mathrm{a}}$ and $\Gamma_{\mathrm{e}}$. Due to thermalization of the phonon bath, these rates satisfy the Kennard-Stepanov relation \cite{Schmitt23,TimBode19}, $\Gamma_{\mathrm{a}}=\Gamma_{\mathrm{e}}e^{\beta \delta}$, at inverse temperature $\beta$, consistent with the Franck-Condon principle. 

\begin{figure}[t]
    \centering
    \includegraphics[width=\linewidth]{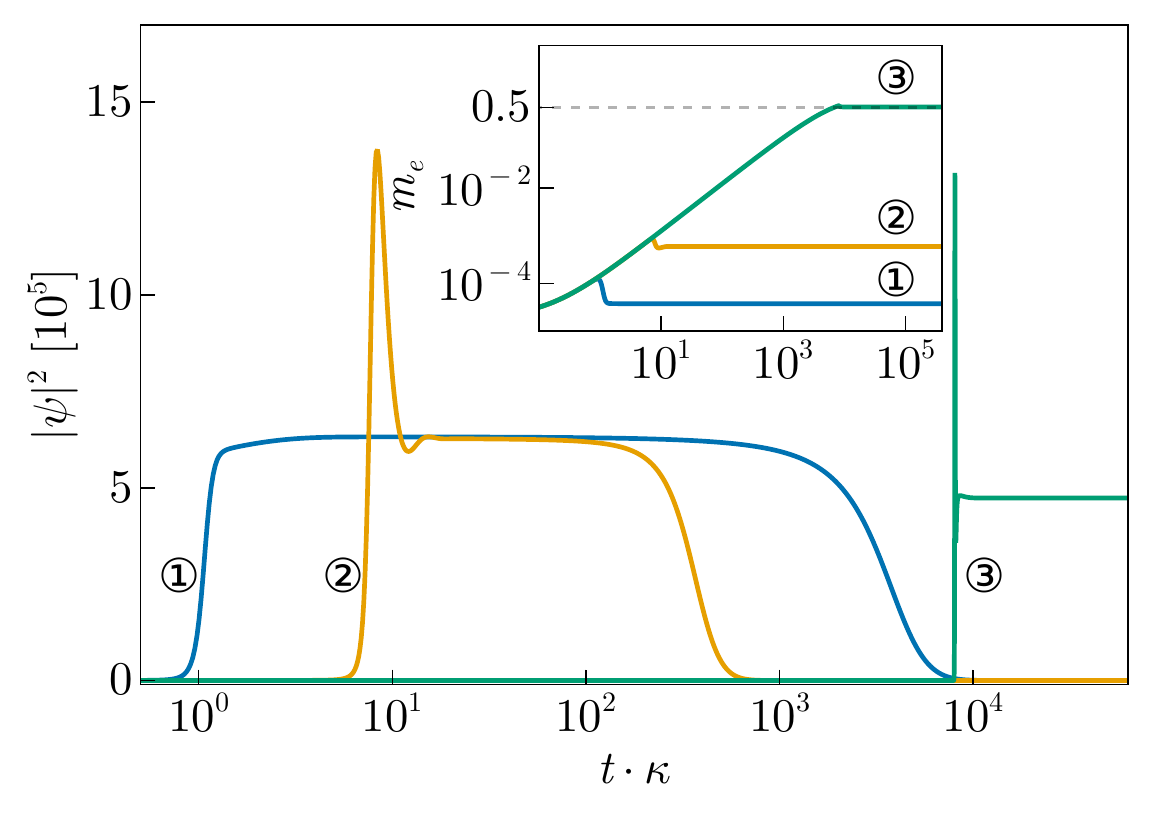}
    \caption{The time evolution of the photon condensate density $|\psi|^2$ is shown for three different values of the absorption and emission ratio, $\log_{10}(\Gamma_{\mathrm{a}}/\Gamma_{\mathrm{e}})=$ \Circled{1} -4.5, \Circled{2} -3.3 and \Circled{3} -0.1, as marked in \cref{fig:stability_map}. Correspondingly, the fraction of excited molecules, $m_{\mathrm{e}}$, is shown in the inset, where the dashed line indicates the lasing threshold, $m_{\mathrm{e}} = 0.5$. \Circled{3} represents a lasing state with infinite lifetime. The system parameters are set to be $\gamma_+/\kappa=1.3\times10^{-4}$, $\gamma_- / \kappa=0$, $M=5.17\times10^9$, $g_\beta/\kappa=10^{-6}$, and $\Gamma_{\mathrm{a}}/\kappa=10^{-9}$ \cite{Kroha21, TimBode19}. Unless stated otherwise, these values are used throughout the paper. The initial conditions for the time evolution were set to an empty cavity and weakly excited molecule bath, $\psi(0)=0, n(0)=0$, $m_{\mathrm{e}}\ll 1$. For consistency, the initial molecule emission amplitude was set to its maximal value allowed by the positive definiteness of the molecular reduced density matrix, $\chi(0)=\chi_{\mathrm{max}}=\sqrt{m_{\mathrm{e}}(0)[1-m_{\mathrm{e}}(0)]} \ll 1$.} 
    \label{fig:time-evol}
\end{figure}
For experimentally relevant molecule numbers in the canonical regime, $M\approx 10^9$ \cite{Weitz10}, direct time evolution of $\hat\rho$ is computationally infeasible, but is facilitated by the fact that quantum coherence between different molecules is negligible due to fast solvent collisions. Therefore, correlation between different molecules can be neglected, allowing their expectation values to factorize. In the regime away from strong photon-molecule (polaron) coupling, we use the nonlinear rate equations for the photon condensate as well as the remaining photonic and molecular expectation values derived in Ref.~\cite{Ray25} from \cref{eq:LME}. These read,
\begin{subequations}
\begin{align}
    \partial_t{\psi} =& -ig_{\beta} M \chi - \left[\kappa - MG_{\psi}(m_{\mathrm{e}})\right]\psi/2   \label{eq:psit}
    \\
    \partial_t{\chi} =& \, i g_{\beta} (2m_{\mathrm{e}}-1) \psi - G_{\chi}(n)\chi/2   \label{eq:chit}
    \\
    \partial_t{n} =& \, 2 g_{\beta} M \mathrm{Im}[\psi^* \chi] - \kappa n + MR(m_{\mathrm{e}},n) \label{eq:nt}
    \\ 
    \partial_t{m}_e =& - 2 g_{\beta} \mathrm{Im}[\psi^* \chi] + \gamma_{+}(1-m_{\mathrm{e}}) - \gamma_{-} m_{\mathrm{e}} \nonumber \\
    & - R(m_{\mathrm{e}},n) \, , \label{eq:met}
\end{align} \label{eq:rate_eqn}%
\end{subequations}% 
where $\psi(t)=\langle\hat{a}\rangle$ and $\chi(t)=\langle\sigma^{-}\rangle$ are the photon condensate amplitude and the molecule de-excitation amplitude, while $n(t)=\langle\hat{a}^{\dagger}\hat{a}\rangle$ and $m_e(t)=(1+\langle\sigma^z\rangle)/2$ represent the total photon number and molecule excitation fraction (per molecule), respectively. $G_{\psi}(m_{\mathrm{e}})$ and $G_{\chi}(n)$ are the dissipation rates for $\psi$ and $\chi$, respectively, and $R(m_{\mathrm{e}},n)$ the photon gain or molecular loss rate through absorption and emission processes,
\begin{subequations}
\begin{align}
G_{\psi}(m_{\mathrm{e}})&=\Gamma_{\mathrm{e}}m_{\mathrm{e}} - \Gamma_{\mathrm{a}}(1-m_{\mathrm{e}}) \label{eq:GPsi}\\  G_{\chi}(n) &=\gamma_{+}+\gamma_{-} + \Gamma_{\mathrm{a}} n + \Gamma_{\mathrm{e}} (n+1) \label{eq:GChi} \\
R(m_{\mathrm{e}},n)&=nG_{\psi}(m_{\mathrm{e}})+\Gamma_{\mathrm{e}}m_{\mathrm{e}}\, . \label{eq:R}
\end{align}
\label{eq:rates}%
\end{subequations}%
The last term in \cref{eq:R} describes spontaneous emission from the molecules. Note that a nonvanishing transition amplitude $\langle\sigma^{-}\rangle\neq 0$ is not related to condensation in the molecule system, but rather means that a single molecule is in a Schr\"odinger cat-like superposition of ground and excited states \cite{Keeling17,Ray25}. Since the spacing of dye molecules is much less than the photon wavelength, $\langle \sigma^{-}_j\rangle =\langle\sigma^{-}\rangle$ are the same on each molecule.    
\begin{figure}[t]
    \centering
    \includegraphics[width=\linewidth]{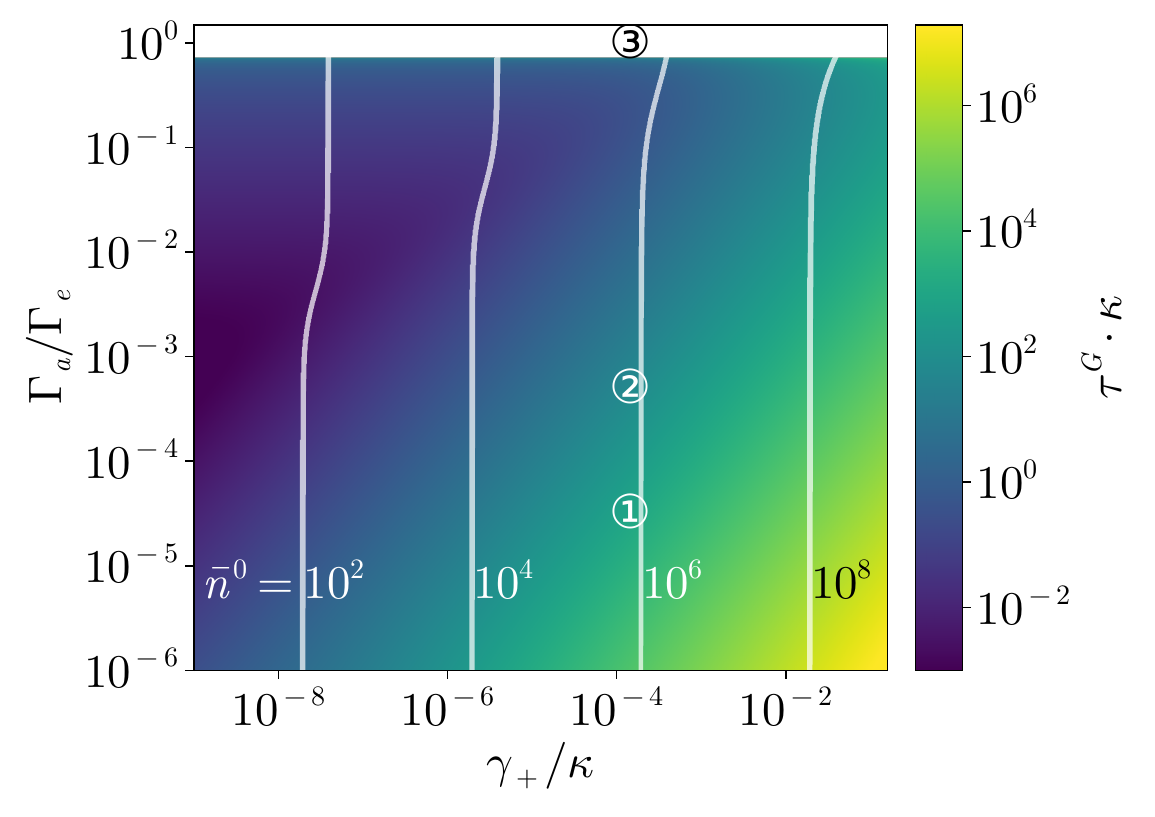}
    \caption{Metastability of the photon condensate. The lifetime $\tau^G$ of the photon condensate is plotted as a color map in the the ratio $\Gamma_{\mathrm{a}}/\Gamma_{\mathrm{e}}$ versus pumping rate $\gamma_+/\kappa$ plane. The vertical, white lines represent constant photon densities $\overline{n}^0$. The white area at the top corresponds to the lasing regime, with infinitely-lived condensate and molecular population inversion, $m_{\mathrm{e}}>1/2$. The condensate time-evolution at the points marked $\Circled{1}$, $\Circled{2}$ and $\Circled{3}$ are shown in \cref{fig:time-evol}.}
    \label{fig:stability_map}
\end{figure}
\begin{figure*}[t]
    \centering
    \begin{overpic}[width=\linewidth]{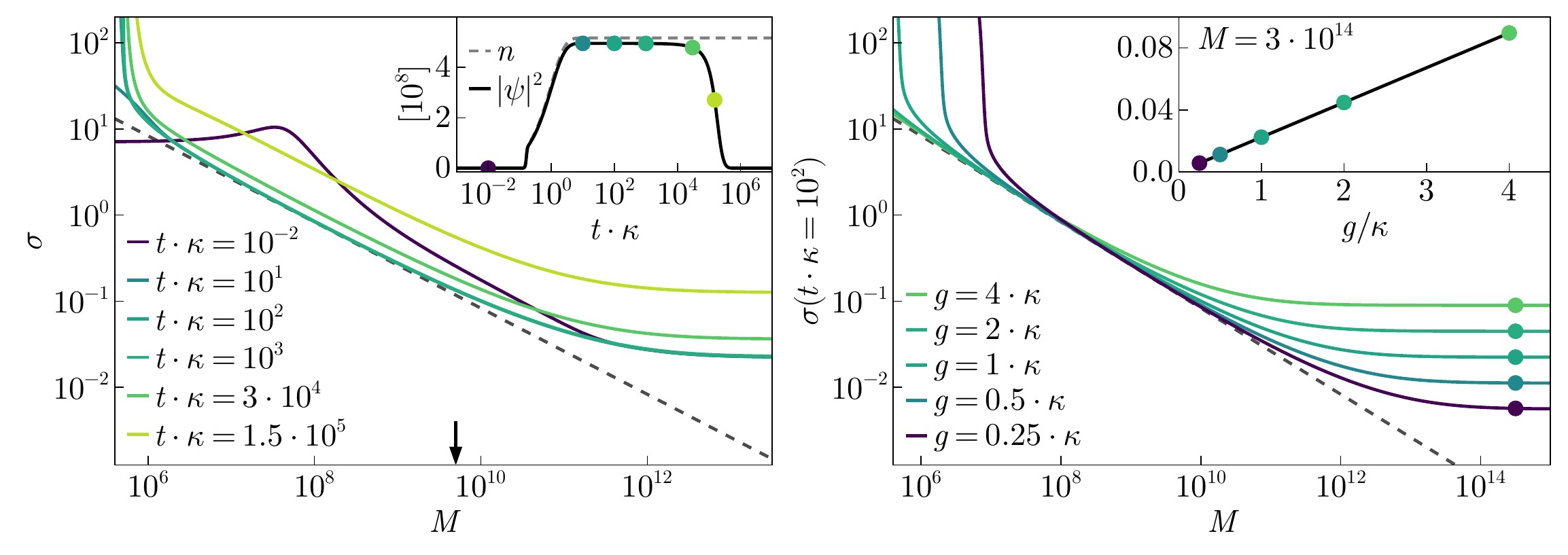}
        \put(12,32.5){\textbf{(a)}}
        \put(65,32.5){\textbf{(b)}} 
    \end{overpic}
    \caption{Order-parameter fluctuations. Normalized fluctuations $\sigma$ of the condensate order parameter $\psi$ are shown as a function of molecule number $M$ for $\gamma_+/\kappa = 0.1$: (a) at different times indicated in the legend and marked in the time evolution (inset); (b) at a plateau time ($t\cdot \kappa=10^2$) for different photon-molecule coupling strengths. The dashed lines in the main plots indicate a slope of $-1/2$, corresponding to the scaling $\sigma \sim 1/\sqrt{M}$. The inset in (a) shows the time evolution of total photon number $n$ and condensate photon number $|\psi|^2$ at $M=5.17\times 10^9$ (arrow). In (b) the effective photon-molecule coupling is controlled by the \emph{bare} Jaynes-Cummings coupling $g$, where $g_\beta \propto g$ and $\Gamma_{\mathrm{a,e}} \propto g^2$ \cite{Pelster18}. The inset shows the saturation values of $\sigma$ versus $g$ in the limit of large system size, $M=3 \times 10^{14}$.}
    \label{fig:th-fluctuation}
\end{figure*}
\section{Fixed points and order-parameter dynamics} \label{sec:dynamics}

Here we briefly recollect the nonlinear dynamics of the model, \cref{eq:rate_eqn,eq:rates}, and present detailed results for the dependence of condensate lifetime, on the cavity detuning $\beta\delta = \ln (\Gamma_a / \Gamma_e)$ and the external pump rate $\gamma_+$. 

Apart from a lasing fixed point (FP) with molecule population inversion $m_{\mathrm{e}} \geq 1/2$, and $\psi(\infty)\neq 0$, the dynamics of the system [\cref{eq:rate_eqn,eq:rates})] exhibits two FPs in the photon-condensate regime ($m_{\mathrm{e}}<1/2$), see \cref{App:FP}. A stable FP $\overline{X}^0$ with vanishing condensate amplitude, $\psi(\infty)=0$ which is always reached in the long-time limit. The second is a so-called ghost FP $\overline{X}^G$, which has a non-vanishing $\psi \neq 0$, but lies outside the physical configuration space since it has an unphysical condensate fraction $\nu = |\psi|^2/n >1$ \cite{Ray25}. All but one of its Lyaponov exponents $\mathrm{Re}(\lambda)$ are attractive (negative) and large (of the order of $\kappa$), while one exponent is small $\mathrm{Re}(\lambda^G)\ll \kappa$ and repulsive (positive), see \cite{Turaev26} and \cref{App:FP} for details. Consequently, the dynamics is quickly attracted toward the ghost FP, with a long-lived plateau in the condensate amplitude $\psi(t)$, since it cannot cross the border to the unphysical realm $\nu>1$. Eventually, the small, positive Lyapunov exponent $\mathrm{Re}(\lambda^G)$ repels the dynamics toward the stabled FP $\overline{X}^0$, where the condensate amplitude ultimately decays to zero. This resulting plateau-like, metastable behavior of the condensate amplitude $\psi(t)$, has a duration typically many orders of magnitude greater than $1/\kappa$. \cref{fig:time-evol} shows the resulting time evolution for representative parameter values given in the figure caption. Detailed results for the condensate lifetime, as extracted analytically from the ghost repulsive Lyapunov exponent as $\tau^G = 1/\mathrm{Re}(\lambda^G)$, are shown in \cref{fig:stability_map} in dependence on the cavity detuning $\beta\delta = \ln (\Gamma_{\mathrm{a}}/ \Gamma_{\mathrm{e}})$, and the external pump rate $\gamma_+$, showing that the condensate lifetime increases with increasing $\gamma_+$ and $\delta$. These lifetimes agree quantitatively well with the plateau duration extracted from the numerical time evolution such as shown in \cref{fig:time-evol} and are consistent with recent experimental findings \cite{Redmann26}. 

\section{Order-parameter fluctuations} \label{sec:fluctuation}

Strikingly, despite the photon BEC being a strongly driven, dissipative system far from equilibrium, experiments show that the condensate obeys a fluctuation-dissipation relation and thus exhibits thermal character \cite{Schmitt23}. So far, we have shown that the driven–dissipative photon condensate is dynamically stabilized. Besides the fluctuation-dissipation theorem, which probes the fluctuation spectrum, the scaling of the fluctuations with system size bears unique information on the thermodynamic behavior of a system. Specifically, for a system in equilibrium, the relative fluctuations of a physical quantity scale inversely with the square root of the system size \cite{Reichl_book16}. To reconcile the analysis of thermal behavior with our single-time rate equation approach, we thus compute the relative fluctuations $\sigma (t)$ of the condensate order parameter at time $t$, 
\begin{equation}
    \sigma (t) = \sqrt{\frac{\langle |\hat{a} - \langle\hat{a}\rangle |^2\rangle }{|\langle\hat{a}\rangle|^2}} = \sqrt{\frac{n(t)-|\psi (t)|^2}{|\psi (t)|^2}}.
\end{equation} 
as a function of the dye molecule number $M$ which plays the role of the system size here, with all other parameters fixed. 
Although $\sigma(t)$ is defined in terms of single-time expectation values, its dynamics is directly connected to the experimentally accessible two-time correlation functions $g^{(1)}(t)$ and $g^{(2)}(t)$, when the reference time, in respect to which the fluctuation is considered, is set to 0. According to the quantum regression theorem, the conditional (with respect to the initial-time value) expectation value entering the first-order correlation function $g^{(1)}(t)$ obeys the same dynamical equations as the condensate amplitude $\psi(t)$, with the initial condition  $\psi(0)$ (see \cref{App:Two-time}). Likewise, the conditional photon-number expectation value entering the second-order correlation function $g^{(2)}(t)$ obeys the same dynamical equations as $n(t)$, with the initial condition $n(0)$ \cite{Kroha21}.

The scaling of $\sigma(t)$ with system size $M$ is shown in \scref{fig:th-fluctuation}{(a)} for different instances of time during the evolution as marked in the inset. It is seen that in the condensate plateau region, the fluctuations indeed exhibit the scaling $\sigma \propto 1/\sqrt{M}$ for an intermediate number of molecules (for our parameters, $5\cdot 10^6\lesssim M\lesssim 5\cdot 10^9$). By contrast, away from the plateau region, the $1/\sqrt{M}$ scaling no longer holds, indicating a nonequilibrium situation. For smaller $M\lesssim 5\cdot 10^6$, $\sigma$ diverges. One can trace this back to the transition to the lasing regime by comparing with the parameter values in the $\delta - \gamma_+$ phase diagram of Ref.~\cite{Ray25}: Reducing the molecule number $M$ at fixed external pump rate $\gamma_+$ drives the molecules to population inversion, $m_{\mathrm{e}}>1/2$. As the lasing transition is approached, the ghost attractor moves further away from the physically accessible region of phase space (see \cref{App:FP}, \cref{fig:lasing-transition}). Consequently the condensate amplitude becomes vanishingly small (cf. phase diagram in Ref.~\cite{Ray25}) resulting in a divergence of $\sigma(t)$. One may view this as critical fluctuations near the transition from the photon condensate to the lasing phase. On the other hand, for larger molecule numbers, $M\gtrsim 10^{12}$, the fluctuations saturate as $M\to\infty$. The origin of these deviations can be extracted from \scref{fig:th-fluctuation}{(b)} which shows the dependence of $\sigma$ on the \emph{bare} Jaynes-Cummings couplings $g$ to the nonequilibrium driven molecule system. With increasing $g$ for a sufficiently large molecule system, $M\gtrsim 10^{12}$, the photon gas thus becomes increasingly influenced by the nonequilibrium molecule dynamics, so that $\sigma$ cannot scale to 0 for $M\to\infty$, but saturates at constant values proportional to $g$, as seen in the inset of \scref{fig:th-fluctuation}{(b)}. By contrast, we find that varying the cavity loss $\kappa$ does not significantly change the saturation values (not shown). This is expected, because $\kappa$ is a coupling to the non-driven environment. 

These combined results provide theoretical evidence that, in the metastable state, the photon condensate exhibits near thermal behavior, with characteristic deviations for small and large systems sizes, thereby accounting for the experimentally observed thermalization \cite{Schmitt14}.
\begin{figure}[t]
    \centering
    \includegraphics[width=\linewidth]{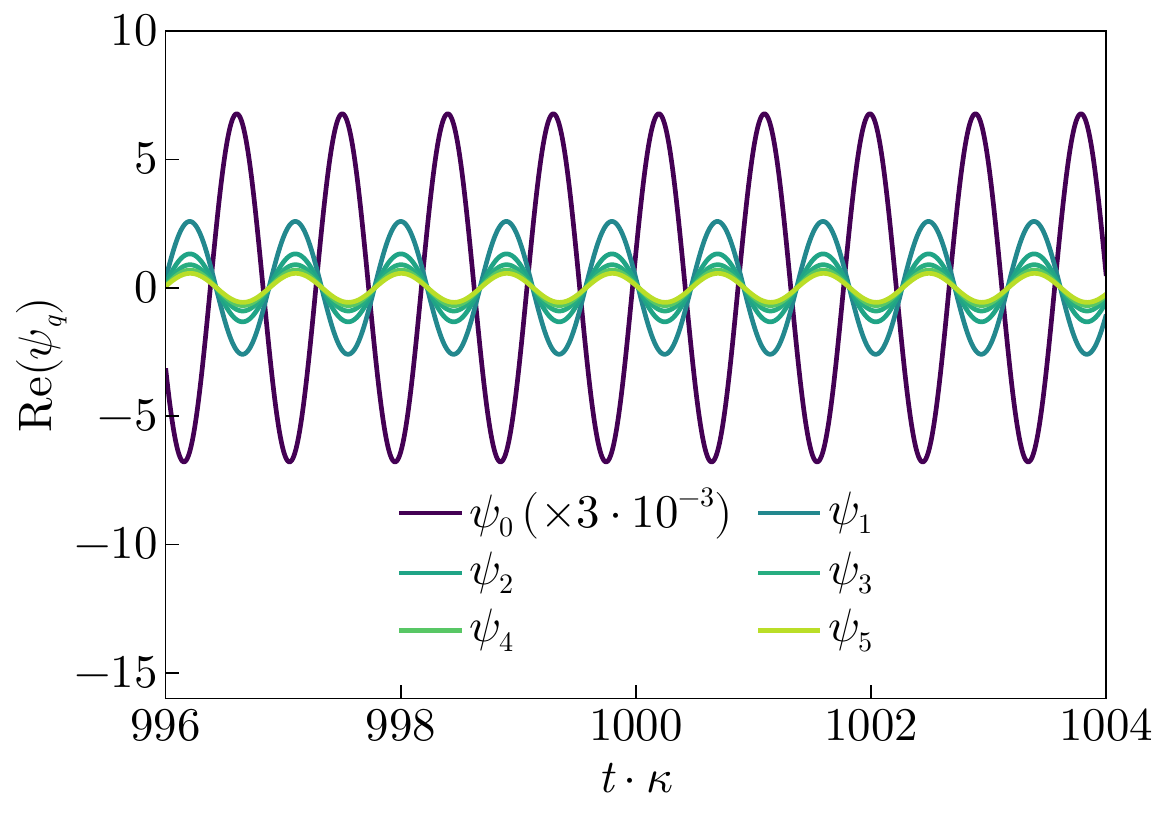}
    \caption{Synchronization of condensate modes. The real parts of the condensate fields, ${\rm Re}(\psi_q)$, for different cavity modes are shown over a zoomed-in time window. Note that these oscillations persist throughout the metastable regime. The ground mode is shown as $\mathrm{Re}(\psi_q)\times (3 \cdot 10^{-3})$ to fit it on the same scale as the other modes. All modes oscillate at a common frequency determined by the condensate energy $-\delta_0$, demonstrating frequency synchronization across the modes. The parameter values are as in \cref{fig:multimode}.} 
    \label{fig:synchronization}
\end{figure}
\begin{figure*}[t]
    \centering
    \begin{overpic}[width=\linewidth]{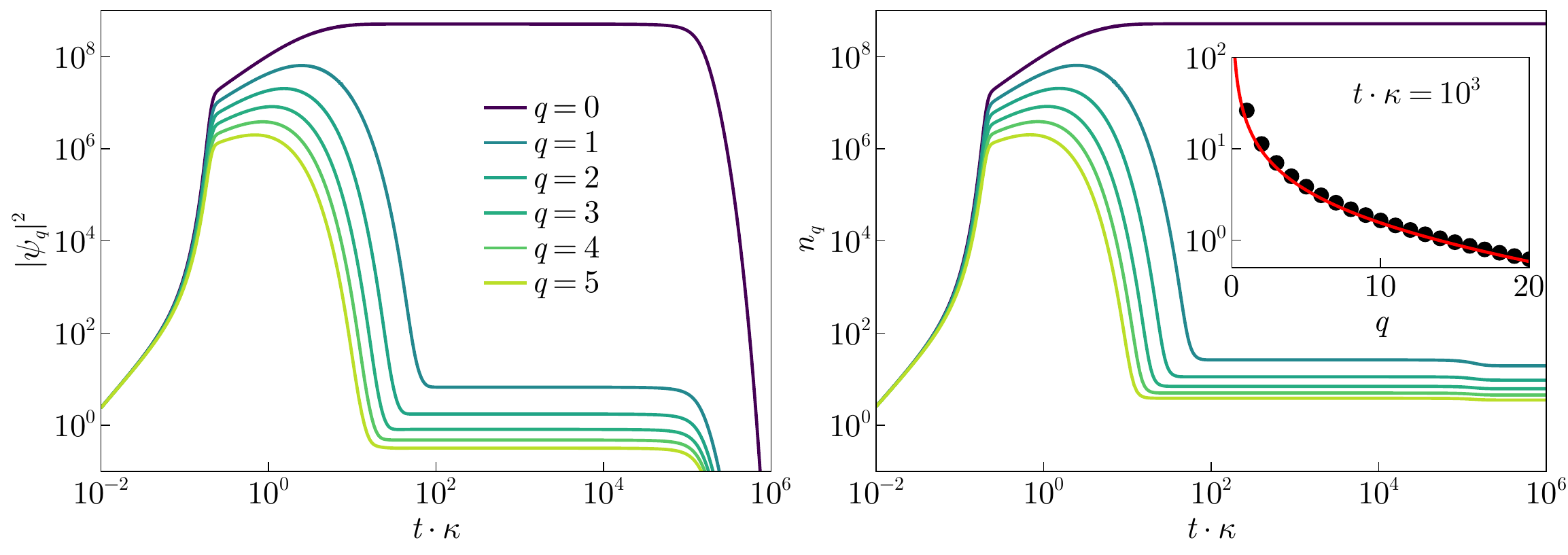}
        \put(8,32){\textbf{(a)}}
        \put(58,32){\textbf{(b)}}
    \end{overpic}
    \caption{Photon thermalization in a multimode cavity. Time evolution of coherent photon number and total photon number, $|\psi_q|^2$ and $n_q$, corresponding to the $q$th mode of a multimode cavity are shown in (a) and (b), respectively for $\gamma_+/\kappa = 0.1$, $M=5.17\times 10^9$, $g_\beta/\kappa=10^{-6}$, and $\Gamma_{\mathrm{a}}/\kappa=10^{-9}$. The numerical simulation is done with $Q=21$ modes, while the dynamics of photon occupancies in the first $6$ modes ($q=0, \cdots, 5$) are displayed. The emission rate for the photonic modes are set as: $\Gamma_{\rm e}^q = \Gamma_{\rm a} e^{-\beta\delta_q}$, where the ground mode detuning is set as, $\beta\delta_0=\beta \delta = -7$ with mode spacing $\beta \omega = 0.05$. The inset in (a) shows the photon number distribution in the cavity modes, $n_q$ vs $q$, at the plateau ($t\cdot \kappa = 10^3$)---the black dots are the numerical data and the solid red line is the Bose-Einstein distribution function.}
    \label{fig:multimode}
\end{figure*}
\section{Dynamics in a multimode cavity and mode synchronization}
\label{sec:multimode}

A more direct characterization of thermalization is possible in multimode photon resonators via the mode-occupation distribution. Thermalization of the photon occupation numbers $n_q$ in the individual modes $q$ has been explained theoretically in the infinite-time limit by Kirton and Keeling \cite{Keeling15}, consistent with experiments \cite{Weitz10,Nyman24a,Pieczarka24}. However, it remains an open question whether a long-lived, metastable condensate plateau persists in a multimode system and whether thermalization occurs in this intermediate state. This question is especially important because in a multimode system additional dephasing channels might degrade the metastable condensate via mode mixing.  

The generalization of the dynamical equations [\cref{eq:rate_eqn,eq:rates}] to multi-mode systems is straight-forward (see \cite{Abouelela26} for more details) and reads,
\begin{subequations}
\begin{align}
    \partial_t{\psi}_q =& -i\delta_q \psi_q -ig_{\beta} M \chi - \left[\kappa - MG_{\psi_q} (m_{\mathrm{e}})\right]\frac{\psi_q}{2}   \label{eq:psit_multi}
    \\
    \partial_t{\chi} =& \, i g_{\beta} (2m_{\mathrm{e}}-1) \sum_{q} d_q \psi_{q}  - G_{\chi}\left(\{n_q\}\right)\frac{\chi}{2}   \label{eq:chit_multi}
    \\
    \partial_t{n}_q =& \, 2 g_{\beta} M \mathrm{Im}[\psi_{q}^* \chi] - \kappa n + MR(m_{\mathrm{e}},\{n_{q}\}) \label{eq:nt_multi}
    \\ 
    \partial_t{m}_e =& - 2 g_{\beta} \sum_{q} d_q \mathrm{Im}[\psi_{q}^* \chi] + \gamma_{+}(1-m_{\mathrm{e}}) - \gamma_{-} m_{\mathrm{e}} \nonumber \\
    & - \sum_{q} d_q R(m_{\mathrm{e}},n_q) \, , \label{eq:met_multi}
\end{align}\label{eq:rate_eqn_multi}%
\end{subequations}%
with the transition rates,
\begin{subequations}
\begin{align}
G_{\psi_q}(m_{\mathrm{e}})&=\Gamma_{\mathrm{e}}^q m_{\mathrm{e}} - \Gamma_{\mathrm{a}}^q (1-m_{\mathrm{e}}) \label{eq:GPsi_multi}\\  
G_{\chi}(\{n_{q}\}) &= \gamma_{+}+\gamma_{-} + \sum_{q} d_q \left[ \Gamma_{\mathrm{a}}^q n_{q} + \Gamma_{\mathrm{e}}^q (n_{q}+1) \right] \label{eq:GChi_multi} \\
R(m_{\mathrm{e}},n_{q})&= n_{q} G_{\psi_q}(m_{\mathrm{e}})+\Gamma_{\mathrm{e}}^q m_{\mathrm{e}}\, . \label{eq:R_multi}
\end{align}\label{eq:rates_multi}%
\end{subequations}%
In the multimode case, each photon mode $q$ is characterized by an individual photon number $n_q$ as well as a $U(1)$ symmetry-breaking amplitude $\psi_q$. Since all photon modes couple equally to the molecule system, they are summed over as shown. For concreteness, we consider a two-dimensional harmonic resonator with equidistant mode frequencies $\Omega_q=\Omega_0+q\,\omega$ and a mode degeneracy of $d_q=q+1$, $q=0,\,1,\,2,\,\dots ,\,Q$. For the numerical computations we take $Q=21$. Note that, in contrast to the single-mode case, the individual mode detunings $\delta_q = \Omega_q-\Delta$ cannot be gauged away, and the Einstein coefficients $\Gamma_{\mathrm{a}}^q$ and $\Gamma_{\mathrm{e}}^q$ also acquire a mode dependence via the Kennard-Stepanov relation $\Gamma_{\mathrm{e}}^q=\Gamma_{\mathrm{a}}^qe^{-\beta \delta_q}$. 

Solving the multimode equations, we first observe that the mutual coupling of the symmetry-breaking amplitudes $\psi_q$ via the molecule system implies that, if one amplitude $\psi_0\neq0$, then all $\psi_q$ must be nonvanishing. Furthermore, since the modes have different detunings $\delta_q$, the question arises if and how a single photon condensate can exist. The numerical solutions of \cref{eq:rate_eqn_multi,eq:rates_multi} show that, after a short transient process of duration $\sim 1/\kappa$, the individual, complex condensate amplitudes $\psi_q$ do oscillate with a common frequency, that is, they synchronize themselves via the coupling to the common bath of molecule excitations $\chi(t)$ and $m_{\mathrm{e}}$. This is seen in \cref{fig:synchronization} which shows stationary, synchronous oscillations of the condensate amplitudes $\psi_q(t)$ in the first six modes in a time window of $[996,1004] 1/\kappa$. This means that a single, quantum-coherent condensate exists with contributions from all modes (even though, for the parameters of \cref{fig:synchronization}, the ground-mode amplitude $\psi_0$ is dominating). This is consistent with the fact that the condensate is not part of the single-particle spectrum, but rather the amplitude of particles \emph{removed} from the system, $\phi_q = \langle \hat{a}_q \rangle$, and agrees with experimental observations \cite{Weitz10,Nyman24a,Pieczarka24}. 

We next analyze the metastability of the multi-mode condensate. In the presence of several modes, it is not feasible to find an analytic solution for an intermediate ghost fixed point like in the single-mode case. Instead we show in \cref{fig:multimode} the numerical time evolution of the condensate densities $|\psi_q|^2$ and the total mode occupation numbers $n_q$ for the first six modes, for the same initial conditions as in \cref{sec:dynamics}. It is seen that also in the multimode case a metastable plateau in the time evolution of all the condensate densities exists, where the densities in excited modes settle at values several orders of magnitude smaller than the ground mode. The initial maxima at a time $\sim 1/\kappa$ are due to initial noneqilibrium pumping by the molecule system. As in the single-mode case the lifetime of the condensate densities is many orders of magnitude larger than $1/\kappa$, before the condensate densities decay to $0$. From the similarity of these time traces to the single-mode case we conjecture that the ghost attractor dynamics is also active in the multimode system. 

Finally, we analyze the mode-occupation distribution function $n_q$ versus $q$ in the plateau region, e.g., at time $\kappa\,t=10^3$. The numerical data (dots) for all 21 modes are shown in the inset of \scref{fig:multimode}{(b)} in comparison to the thermal Bose-Einstein distribution $b(q)= 1/(e^{\beta\delta_q}-1)$ (red line), with vanishing chemical potential in the condensate phase and inverse temperature $\beta$ fixed by the ambient temperature of the molecular phonon bath. Note that since the Bose-Einstein distribution was not fitted, the excellent agreement indicates that the multimode photon system is in near thermodynamic equilibrium not only in the infinite-time limit but also in the metastable plateau region of the condensate. 

\section{Exceptional points and non-Hermitian Phase transitions} \label{sec:NHPandEP}

Having demonstrated thermal behavior of the metastable, but long-lived photon condensate in both single- and multimode cavities, we now turn to the temporal relaxation into and out of this state, which nevertheless are genuine nonequilibrium processes. In particular, a NHPT  of the photon-density correlation function $g^{(2)}(t)$ has been observed experimentally \cite{Kroha21} and analyzed theoretically \cite{Kroha21,TimBode19} within the BEC phase of a driven-dissipative photon gas. This occurs when the condensed system is driven out of its stationary state either by a random fluctuation or by an initial pump pulse. Since the photon condensate amplitude $\psi(t)$ is metastable, a NHPT with an EP might occur during the relaxation into the metastable state (toward the ghost FP $\overline{X}^G$), out of this state, as well as into the final stationary state (the stable FP $\overline{X}^0$). Access to $\psi(t)$ can be provided by ongoing interferometric experiments \cite{Redmann26} via the first-order (amplitude) correlation function $g^{(1)}(t)$. 

At an EP two eigenfrequencies of the dissipative, two-component photon-molecule system become degenerate, and the two eigenvectors align \cite{Kroha21}. We restrict ourselves here to a single-mode cavity, where analytic calculations are feasible (see below). We thus linearize the dynamics of \cref{eq:rate_eqn} about both fixed points (FPs) and compute the corresponding stability matrices $\mathcal{M}^0$ and $\mathcal{M}^G$. We analyze their eigenvalues $\lambda^G$ and $\lambda^0$ in dependence on the pump rate $\gamma_+$, or equivalently the steady-state photon number $\overline{n}^0$.

\begin{figure}[t]
    \centering
    \begin{overpic}[width=\linewidth]{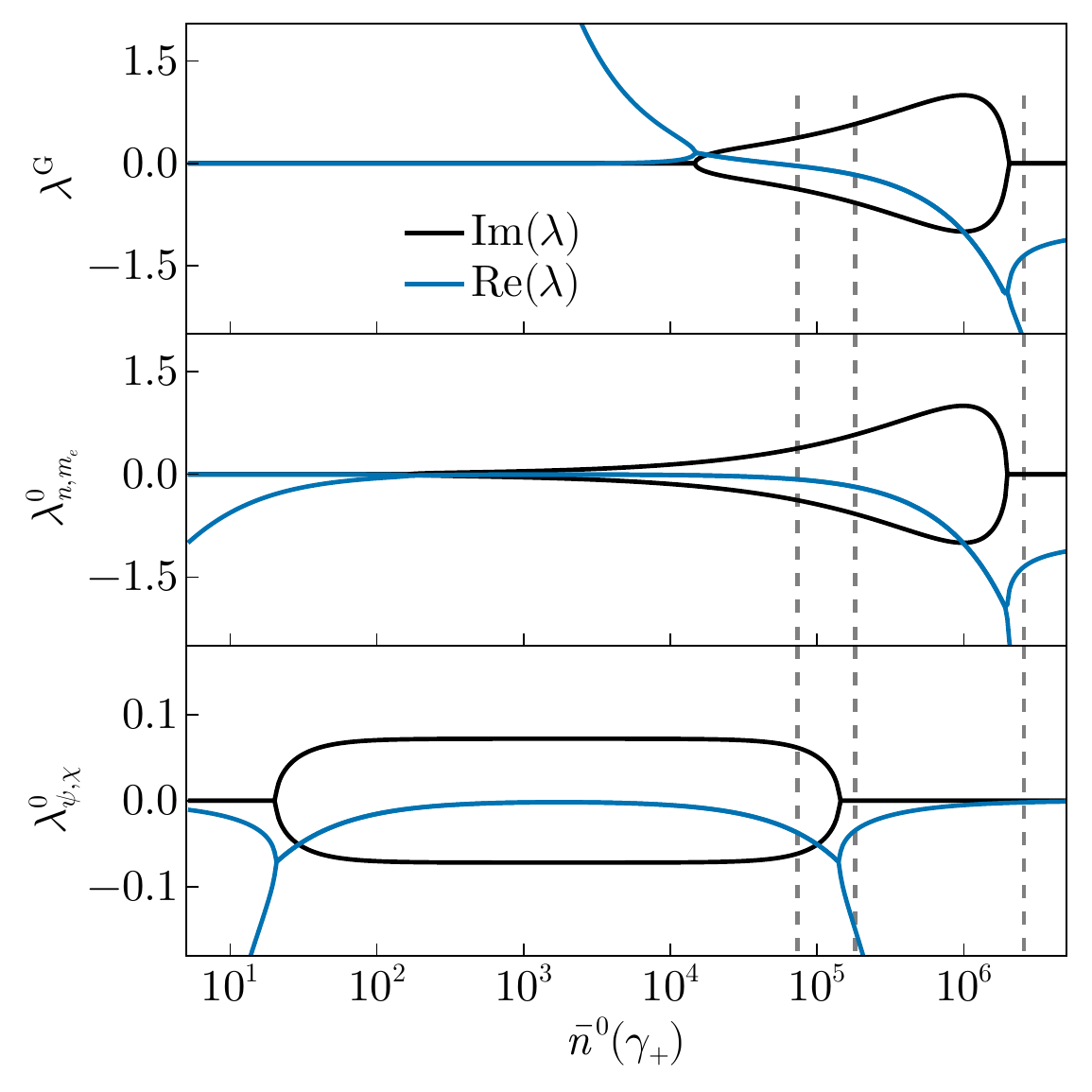}
        \put(19,93){\textbf{(a)}}
        \put(19,64.6){\textbf{(b)}}
        \put(19,36){\textbf{(c)}}
        \put(70.7,93){\Circled{iii}}
        \put(76.3,93){\Circled{ii}}
        \put(91.7,93){\Circled{i}}
    \end{overpic}
    \caption{Exceptional points. Real (right axis) and imaginary (left axis) parts of the eigenvalues $\lambda$ of the stability matrix $\mathcal{M}$ as a function of the steady-state photon number $\overline{n}^0$. (a) represents the two eigenvalues involved in the EP degeneracy for $\mathcal{M}^{\mathrm{G}}$. (b) and (c) show the eigenvalues $\lambda^0_{n,m_{\mathrm{e}}}$ in the number-density sector and $\lambda^0_{\psi,\chi}$ in the condensate-amplitude sector of $\mathcal{M}^0$, respectively. The eigenvalues undergo NHPTs at their corresponding EPs as the pump rate $\gamma_{+}/\kappa$, and thereby $\overline{n}^0$, is varied. Dashed vertical lines mark the pump rates at which the relaxation dynamics are shown in \cref{fig:dynamics-EP}.}
    \label{fig:EP}
\end{figure}

While in the metastable regime, the total-density modes and the amplitude modes are coupled, the stability matrix $\mathcal{M}^0$ of $\overline{X}^0$ assumes a block-diagonal form (see \cref{App:FP}, for $\psi,\chi=0$),
\begin{equation}
    \mathcal{M}^0=\mathcal{M}^0_{n,m_{\mathrm{e}}} \oplus \mathcal{M}^0_{\psi,\chi} \oplus \mathcal{M}^0_{\psi^*,\chi^*},
\end{equation}
such that the linearized dynamics decouple into the density sector, $\mathcal{M}^0_{n,m_{\mathrm{e}}}$, and the complex amplitude sectors, $\mathcal{M}_{\psi,\chi}$, $\mathcal{M}_{\psi^*,\chi^*}$. 
\begin{figure}[t]
    \centering
    \begin{overpic}[width=\linewidth]{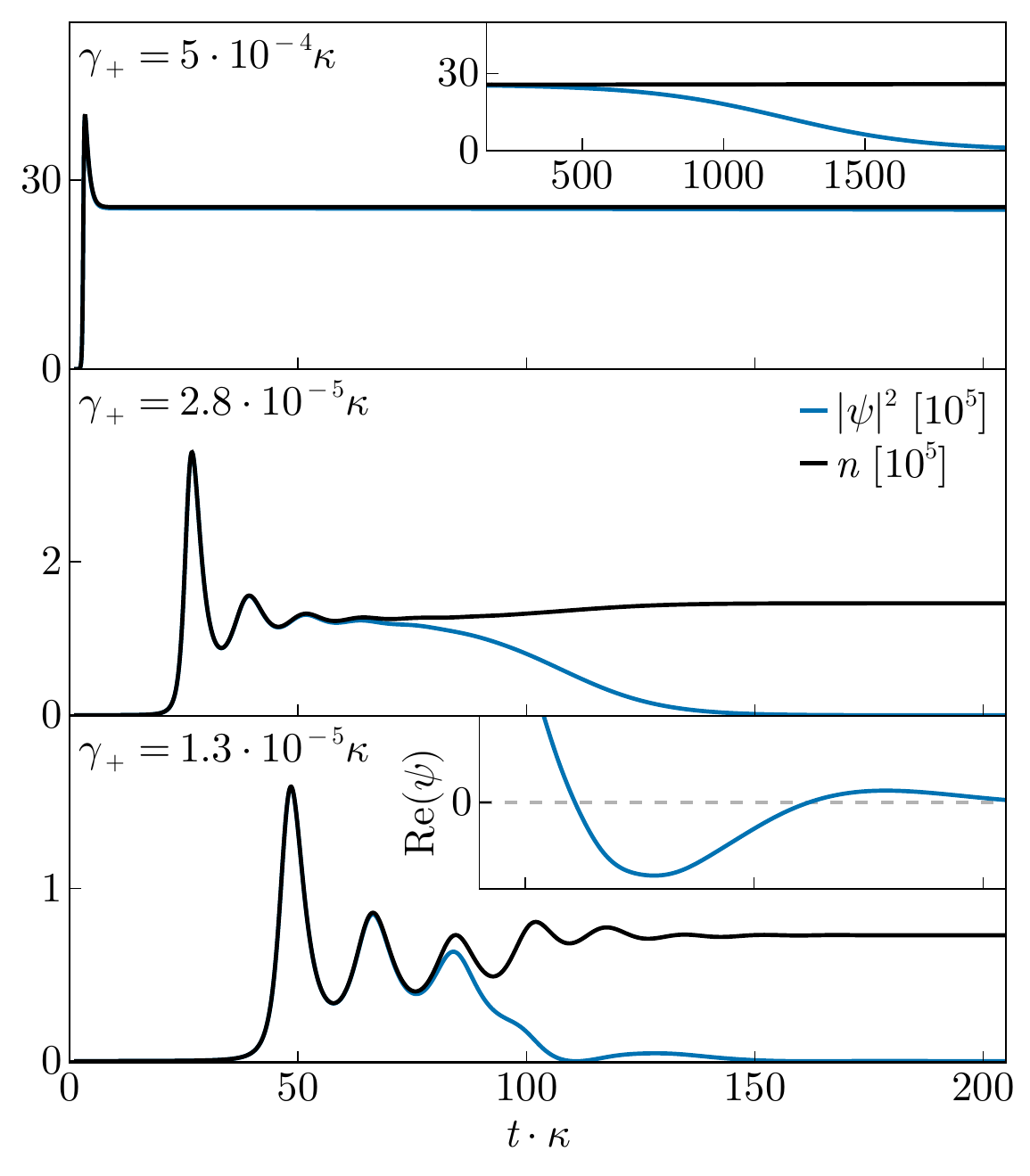}
        \put(0.35,94){\Circled{i}}
        \put(0.35,64){\Circled{ii}}
        \put(0.35,34){\Circled{iii}}
    \end{overpic}
    \caption{Biexponential and oscillatory relaxation. Time evolutions of total photon number $n(t)$ (black) and condensate density $|\psi(t)|^2$ (blue), undergoing oscillatory or biexponential relaxation dynamics, are shown for different pump rates $\gamma_{+}/\kappa$, as indicated by \Circled{i}, \Circled{ii} and \Circled{iii}, respectively, in \cref{fig:EP}. The inset in (a) shows biexoponential relaxation, while the inset in (c) shows multiple phase jumps in $\mathrm{Re}(\psi)$ associated with zeroes of the oscillatory relaxation of $|\psi|^2$ toward the steady-state FP $\overline{X}^0$.}
    \label{fig:dynamics-EP}
\end{figure}
As seen in \cref{fig:EP}, $\lambda^{0}_{n,m_{\mathrm{e}}}$, $\lambda^{0}_{\psi,\chi}$, and $\lambda^{G}$  exhibit EPs separating regimes with two real eigenvalues from regimes with two complex conjugate pairs, corresponding to biexponential and oscillatory relaxation behavior, respectively. Interestingly, in the relaxation to the long-time stable regime NHPTs of the density modes $(n,m_{\mathrm{e}})$ and of the amplitude modes $(\psi,\chi)$ occur at different photon numbers $\overline{n}^0$. This indicates that the condensate undergoes a qualitatively different dynamical transition than the photon number. 

The interplay between the two FPs gives rise to a variety of NHPTs in the relaxation toward the metastable and the long-time stable FPs, $\overline{X}^{\mathrm{G}}$ and $\overline{X}^{0}$, respectively. This is illustrated by the time evolution of the total photon number $n$ and the condensate amplitude $\psi$ shown in \cref{fig:dynamics-EP} for three representative values of the pump rate $\gamma_+$, marked by the dashed lines \Circled{i}, \Circled{ii}, and \Circled{iii} in \cref{fig:EP}. \Circled{i} shows non-oscillatory, biexponential relaxation of both $|\psi(t)|^2$ and $n(t)$ toward $\overline{X}^{\mathrm{G}}$ as well as $\overline{X}^{0}$. {\Circled{ii}} displays oscillatory relaxation toward $\overline{X}^{\mathrm{G}}$, but biexponential relaxation toward $\overline{X}^{0}$, \Circled{iii} shows oscillatory relaxation toward $\overline{X}^{\mathrm{G}}$ as well as $\overline{X}^{0}$ at different oscillation frequencies.

\section{Conclusion} \label{sec:conclusion}

In summary, We have analyzed the properties of the long-lived, yet metastable photon Bose-Einstein condensate system in an optical microsesonator, coupled to an externally driven system of dye-molecule, comprised of electronic and vibrational excitations. 

In the metastable phase, the system exhibits thermal behavior. The BEC order-parameter fluctuations $\sigma$ scale as the inverse square root of the system size (i.e., molecule number $M$) as in thermodynamic equilibrium. In multimode system, the photon number distribution across the modes agrees excellently with a Bose-Einstein distribution without parameter fitting, with vanishing chemical potential and an ambient temperature set by the dye reservoir. The metastable condensate dynamics reveals another property of multi-mode condensates: if there is a condensate amplitude in one mode, the coupling via the molecule excitations implies $U(1)$-symmetry-breaking amplitudes in all modes. However, these complex amplitudes do not oscillate with the individual mode frequencies, but are synchronized via the single molecule-excitation density. This is consistent with the fact that the condensate amplitude is a many-body expectation value and not part of the single-photon resonator spectrum. 
Deviations from thermal behavior occur for both small and large $M$ due to the proximity to a lasing transition and the influence of nonequilibrium driving processes in the molecule bath, respectively. This quasithermal behavior is important for experiments on thermodynamics of open photon condensates. 

By contrast, the relaxation dynamics from an initially pumped state into the metastable condensate and, at long times, into the stable photon gas are genuinely nonequilibrium in character. In particular, they can exhibit multiple non-Hermitian phase transitions. We analyzed the complex interplay of these transitions characterized by changes between biexponential and oscillatory relaxation. These signatures should be observable in experiments measuring both, the temporal photon-density correlation function $g^{(2)}(t)$ and the condensate-amplitude correlation function $g^{(1)}(t)$.  

\section*{Acknowledgments}
We thank Michael Kajan, Jonathan Keeling, Andreas Redmann, Julian Schmitt, Frank Vewinger, and Martin Weitz for insightful discussions. S.R. acknowledges RPTU University Kaiserslautern-Landau, where part of this work was carried out. This work was funded by the Deutsche Forschungsgemeinschaft (DFG) under Germany's Excellence Strategy-Cluster of Excellence Matter and Light for Quantum Computing, ML4Q (No. 390534769) and through the DFG Collaborative Research Center CRC 185 OSCAR (No. 277625399). 

Data availability -- The data that support the findings of this article are openly available \cite{Zenodo-data}. 

\vspace*{0.2cm}

\appendix

\section{Fixed points and their stability}
\label{App:FP}

For completeness, we briefly outline the solutions of the fixed points (FPs) $\overline{X}=(\overline{\psi},\overline{\chi},\overline{m}_e,\overline{n})$ of the dynamical equations Eq.~\eqref{eq:rate_eqn} by setting the time derivatives to zero and solving the resulting set of equations. Apart from a solution $\overline{X}^0$ with vanishing symmetry-breaking quantities $\overline{\psi}$ and $\overline{\chi}$, which always exists, for the nonvanishing solutions we obtain the relation
\begin{equation}
4 g_{\beta}^2 M (2\overline{m}_e-1) - G{\chi}(\overline{n}) \left[\kappa-MG_{\psi}(\overline{m}_e)\right] = 0 .
\label{eq:determinant}
\end{equation}
Using \cref{eq:met,eq:nt}, we derive a relation between the fraction of molecule excitation $\overline{m}_e$ and photon number $\overline{n}$,
\begin{equation}
\overline{m}_e = \left( \gamma_+ - \frac{\kappa \overline{n}}{M} \right)/(\gamma_+ + \gamma_-),
\label{eq:m_e-n}
\end{equation}
with which \cref{eq:determinant} becomes a quadratic equation in $\overline{n}$. Additionally, the expression for the condensate fraction reads as
\begin{equation}
\overline{\nu} =
1-\frac{\Gamma_e \overline{m}_e G_{\chi}(\overline{n})}{4 g_{\beta}^2 (2\overline{m}_e-1)} .
\label{eq:nu}
\end{equation}
Since $G_{\chi}(\overline{n})>0$, the only way to get a physical solution with a finite condensate is when $\overline{m}_e > 1/2$, corresponding to molecular population inversion, which we define as the lasing FP $\overline{X}^{\mathrm{L}}$. In contrast, for $\overline{m}_e < 1/2$, the finite condensate FP $\overline{X}^{\mathrm{G}}$ is unphysical with $\overline{\nu}^{\mathrm{G}} > 1$.

To understand the structure of the FPs in different dynamical regimes, we perform a stability analysis of the above FPs. This is done by defining the stability matrix $\mathcal{M}_{ij} = \partial f_i / \partial x_j |_{\overline{X}} $, where $f_i$ corresponds to the right hand side of \cref{eq:rate_eqn} and $x_j \in X$, obtaining
\begin{widetext}
    \begin{equation*}
    \mathcal{M}=
        \begin{pmatrix}
            -[\kappa - MG_{\psi}(\overline{m}_e)]/2 & 0 & -i g_\beta M & 0 & (\Gamma_{\mathrm{a}} +\Gamma_{\mathrm{e}}) M \overline{\psi}/2 & 0 \\
            0 & -[\kappa - MG_{\psi}(\overline{m}_e)]/2 & 0 & i g_\beta M & (\Gamma_{\mathrm{a}} +\Gamma_{\mathrm{e}}) M \overline{\psi}^* / 2 & 0 \\
            i g_\beta (2 \overline{m}_e-1) & 0 & -G_{\chi}(\overline{n})/2 & 0 & i 2 g_\beta \overline{\psi} & -(\Gamma_{\mathrm{a}}+\Gamma_{\mathrm{e}}) \overline{\chi}/2 \\
            0 & -i g_\beta (2 \overline{m}_e-1) & 0 & -G_{\chi}(\overline{n})/2 & -i 2 g_\beta \overline{\psi}^* & -(\Gamma_{\mathrm{a}}+\Gamma_{\mathrm{e}}) \overline{\chi}^* / 2 \\
            -i g_\beta \overline{\chi}^* & i g_\beta \overline{\chi} & i g_\beta \overline{\psi}^* & - i g_\beta \overline{\psi} & -G_{\chi}(\overline{n}) & -G_{\psi}(\overline{m}_e) \\
            i g_\beta M \overline{\chi}^* & -i g_\beta M\overline{\chi} & -i g_\beta M\overline{\psi}^* &  i g_\beta M \overline{\psi} & M\left[\Gamma_{\mathrm{a}}\overline{n}+\Gamma_{\mathrm{e}}(\overline{n}+1) \right] & -[\kappa - MG_{\psi}(\overline{m}_e)]
        \end{pmatrix}.
    \end{equation*}
\end{widetext}
Note that the complex conjugate equations and variables were taken into account. Performing the stability analysis for $\overline{X}^{\mathrm{G}}$ reveals its hyperbolic nature, i.e. all eigenvalues have negative real parts except for at least one with a very small positive Lyapunov exponent. This corresponds to a long characteristic time scale, $\tau^G=1/\mathrm{Re}(\lambda^G) \approx 300 ~\kappa^{-1}$, resulting in an extremely slow repelling dynamics from $\overline{X}^{\mathrm{G}}$ and the emergence of the condensate plateau shown by \Circled{1} in \cref{fig:time-evol}.

\begin{figure}[t]
    \centering
    \includegraphics[width=\linewidth]{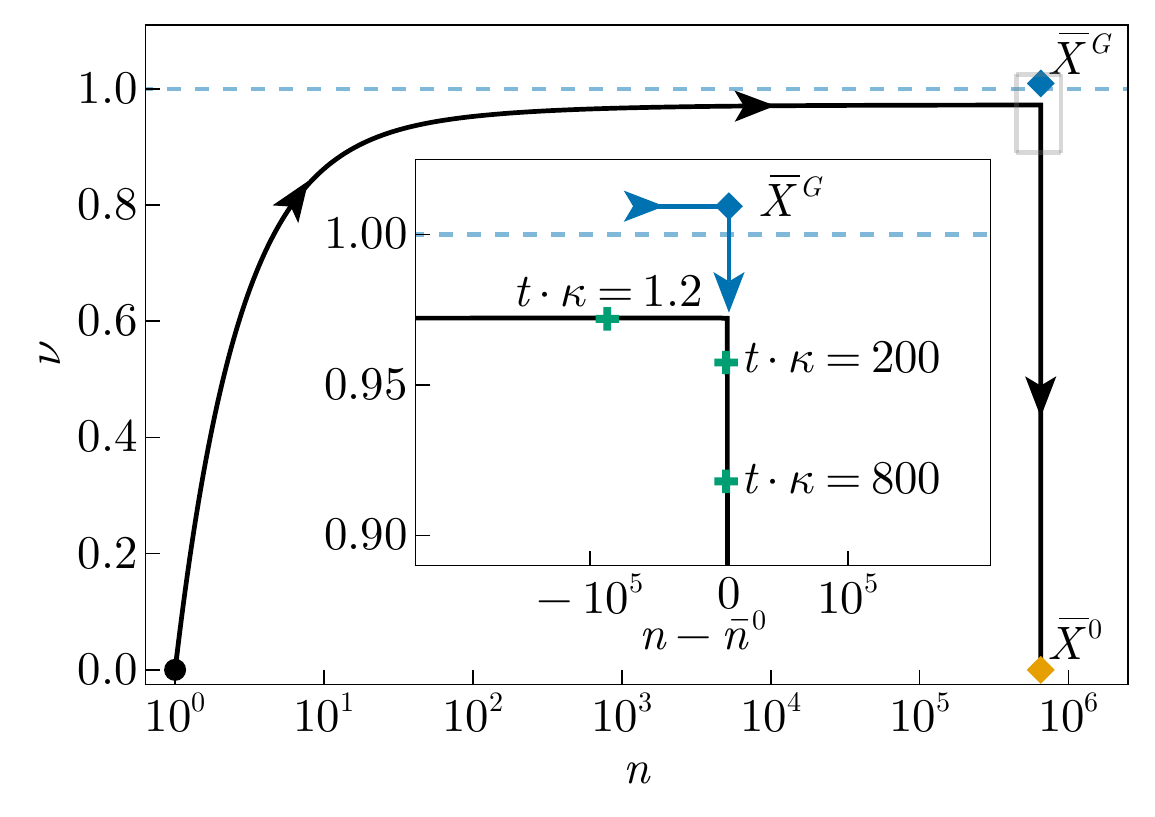}
    \caption{Flow diagram. The projection of the phase-space trajectory corresponding to the time trace \Circled{1} of \cref{fig:time-evol} is shown in the plane of condensate fraction $\nu$ versus total photon number $n$. The slow dynamics near the unphysical FP $\overline{X}^{\mathrm{G}}$ (blue diamond) are demonstrated in the inset by the times of the evolution. The dashed line represents the physical boundary of $\nu=1$. The incoming and outgoing arrows from $\overline{X}^{\mathrm{G}}$ indicate the slowest attractive and repulsive eigenmodes. $\overline{X}^0$ (orange diamond) denotes the steady-state FP with photon number $\overline{n}^0$. 
    }
    \label{fig:ghost}
\end{figure}

\begin{figure}
    \centering
        \begin{overpic}[width=\linewidth]{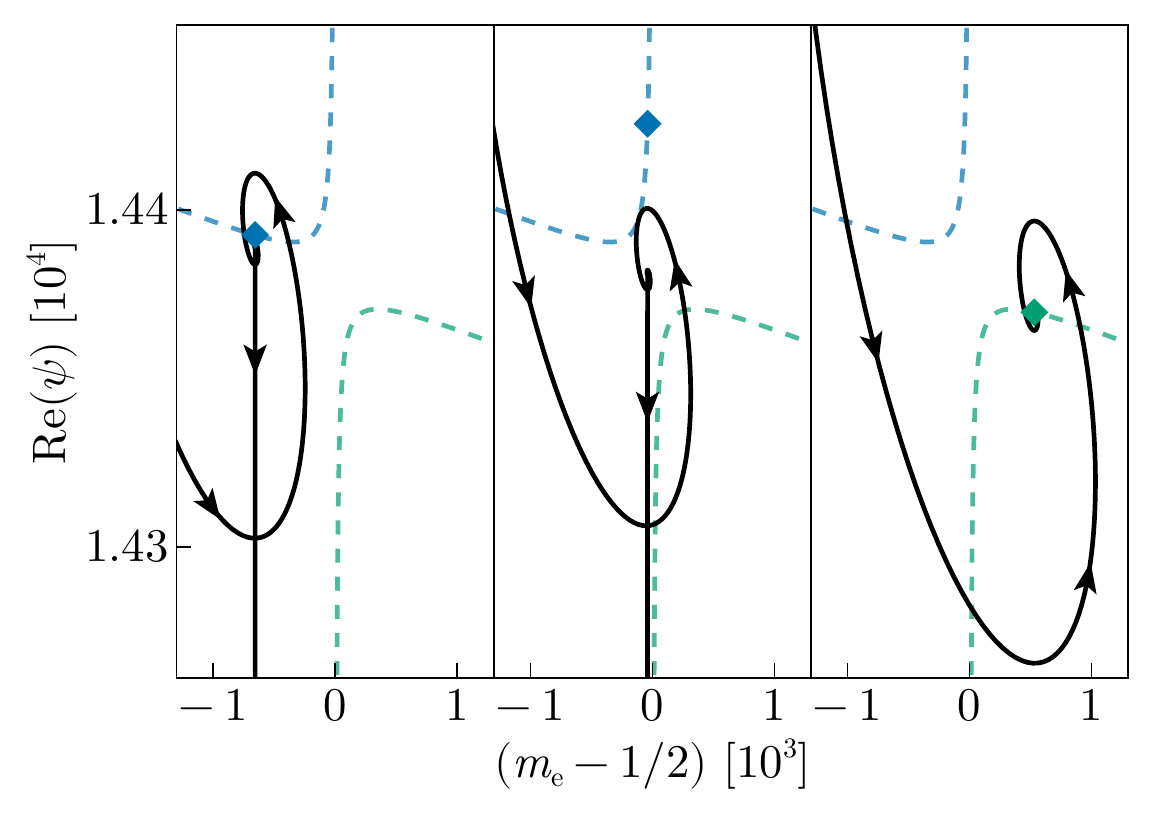}
        \put(18,62){\textbf{(a)}}
        \put(45,62){\textbf{(b)}}
        \put(75,62){\textbf{(c)}}        
    \end{overpic}
    \caption{ Transition to laser. Projections of trajectories on to $m_{\mathrm{e}}$ versus $\mathrm{Re}(\psi)$ plane are shown in (a-c) for different ratios of $\Gamma_{\mathrm{a}}/\Gamma_{\mathrm{e}}$ very close to the laser boundary (see \cref{fig:stability_map}). While (a-b) corresponds to the stabilized condensate regime, with unphysical the FPs $\overline{X}^{\mathrm{G}}$ denoted by the blue diamonds, (c) denotes the lasing regime, with $\overline{X}^{\mathrm{L}}$ denoted by the green diamond. For reference, the blue and green dashed lines show the positions of the FPs, obtained by continuously varying $\Gamma_{\mathrm{a}}/\Gamma_{\mathrm{e}}$ in the same plot.}
    \label{fig:lasing-transition}
\end{figure}

In the lasing regime ($\Gamma_{\mathrm{a}}/\Gamma_{\mathrm{e}} \sim 1$), the FP $\overline{X}^0$ becomes unstable, while $\overline{X}^{\mathrm{L}}$ emerges as a stable FP, indicating a transition from the condensate regime to laser.
This is illustrated in \cref{fig:lasing-transition}, where projections of $\overline{X}^{\mathrm{G}}$ and $\overline{X}^{\mathrm{L}}$ are plotted in the $\mathrm{Re}(\psi)$–$m_{\mathrm{e}}$ plane as the ratio $\Gamma_{\mathrm{a}}/\Gamma_{\mathrm{e}}$ is varied. The trajectories in \scref{fig:lasing-transition}{(a,b)} exhibit the characteristic condensate dynamics stabilized by $\overline{X}^{\mathrm{G}}$ (see also \cref{fig:ghost}). In contrast, in the lasing regime, the dynamics is attracted to and relaxes toward the stable $\overline{X}^{\mathrm{L}}$ with finite condensate amplitude and population inversion, as shown in \scref{fig:lasing-transition}{(c)}.

\FloatBarrier

\section{Two-time correlation functions}
\label{App:Two-time}

In this Appendix, we establish the relation between the experimentally accessible two-time correlation functions $g^{(1)}(\tau)$ and $g^{(2)}(\tau)$ and the single-time expectation values obtained from the rate equations discussed in the main text. This relation follows from the quantum regression theorem, which shows that the same Liouvillian governing the single-time expectation values also determines the dynamics of stationary two-time correlation functions.

The experimentally measured second-order photon correlation function is defined as
\begin{equation}
    g^{(2)}(\tau) = \left. \frac{\langle \hat{n}(t+\tau)\hat{n}(t) \rangle}{\langle \hat{n}(t) \rangle^2} \right|_{t\rightarrow\infty},
\end{equation}
where $\hat{n}=\hat{a}^\dagger\hat{a}$.
Using the explicit time evolution generated by the Lindblad master equation, this expression can be written as
\begin{equation}
    g^{(2)}(\tau) = \frac{{\rm Tr} \!\left[ \hat{a}^\dagger\hat{a}\, e^{\hat{\mathcal{L}}\tau} \hat{a}^\dagger\hat{a}\, \hat{\rho}_\infty \right]}{\left[{\rm Tr}(\hat{a}^\dagger\hat{a}\, \hat{\rho}_\infty) \right]^2} = \frac{{\rm Tr}\!\left[\hat{a}^\dagger\hat{a}\, e^{\hat{\mathcal{L}}\tau} \hat{\rho}^{(\hat{a}^\dagger\hat{a})}_\infty \right]}{\left(\bar n^{0}\right)^2},
\end{equation}
where
\begin{equation}
    \hat{\rho}^{(\hat{a}^\dagger\hat{a})}_\infty = \hat{a}^\dagger\hat{a}\, \hat{\rho}_\infty,
\end{equation}
and
\(\hat{\rho}_\infty = \lim_{t\rightarrow\infty} \hat{\rho}(t) \) is the steady-state density matrix.

More generally, for a stationary two-time correlation function involving operators $\hat{A}$ and $\hat{B}$, we define
\begin{equation}
    \langle \hat{A} \rangle^{(\hat{B})}(\tau) = {\rm Tr} \!\left[\hat{A}\, e^{\hat{\mathcal{L}}\tau} \hat{\rho}^{(\hat{B})}_\infty \right], \qquad
    \hat{\rho}^{(\hat{B})}_\infty = \hat{B}\hat{\rho}_\infty.
    \label{eq:two-time_not}
\end{equation}
The quantity $\langle\hat{A}\rangle^{(\hat{B})}$ may therefore be interpreted as the expectation value of $\hat{A}$ with respect to the conditioned density matrix $\hat{\rho}^{(\hat{B})}_\infty$.

Introducing the conditioned photon number expectation value
\begin{equation}
    n^{(\hat{a}^\dagger\hat{a})}(\tau) = \langle \hat{a}^\dagger\hat{a} \rangle^{(\hat{a}^\dagger\hat{a})},
\end{equation}
the second-order correlation function becomes
\begin{equation}
    g^{(2)}(\tau) = \frac{n^{(\hat{a}^\dagger\hat{a})}(\tau)}{(\bar n^0)^2}.
\end{equation}

The first-order correlation function is treated analogously,
\begin{align}
    g^{(1)}(\tau) &= \left. \frac{\langle\hat{a}(t+\tau)\hat{a}^\dagger(t) \rangle}{\langle\hat{a}^\dagger(t)\hat{a}(t)\rangle}\right|_{t\rightarrow\infty} \nonumber\\
    &= \frac{{\rm Tr}\!\left[\hat{a}\, e^{\hat{\mathcal{L}}\tau}\hat{a}^\dagger\hat{\rho}_\infty\right]}{\bar n^0}
    = \frac{{\rm Tr}\!\left[\hat{a}\, e^{\hat{\mathcal{L}}\tau}\hat{\rho}^{(\hat{a}^\dagger)}_\infty \right]}{\bar n^0},
\end{align}
where
\begin{equation}
    \hat{\rho}^{(\hat{a}^\dagger)}_\infty = \hat{a}^\dagger \hat{\rho}_\infty,
\end{equation}
and we define
\begin{equation}
    \psi^{(\hat{a}^\dagger)}(\tau) = \langle \hat{a} \rangle^{(\hat{a}^\dagger)} = {\rm Tr}\!\left[\hat{a}\, e^{\hat{\mathcal{L}}\tau}\hat{\rho}^{(\hat{a}^\dagger)}_\infty \right].
\end{equation}

According to the quantum regression theorem \cite{Petruccione_book, Kroha21}, for a Markovian open quantum system obeying the Lindblad master equation
\begin{equation}
    \frac{d\hat{\rho}}{dt} = \hat{\mathcal{L}} \hat{\rho},
\end{equation}
the evolution of stationary two-time correlation functions is generated by the same Liouvillian superoperator that governs the evolution of single-time expectation values. In the present notation this follows directly from
\begin{equation}
    \frac{d}{d\tau}\hat{\rho}^{(\hat{B})}_\infty(\tau) = \frac{d}{d\tau} \left(e^{\hat{\mathcal{L}}\tau}\hat{\rho}^{(\hat{B})}_\infty \right) = \hat{\mathcal{L}}\hat{\rho}^{(\hat{B})}_\infty(\tau),
\end{equation}
where only the initial conditioned density matrix differs from the single-time evolution.

Consequently, the conditioned expectation values $\psi^{(\hat{a}^\dagger)}(\tau)$ and $n^{(\hat{a}^\dagger\hat{a})}(\tau)$ satisfy the same dynamical equations as the corresponding single-time expectation values $\psi(\tau)$ and $n(\tau)$, with different initial conditions determined by $\hat{\rho}^{(\hat{a}^\dagger)}_\infty$ and $\hat{\rho}^{(\hat{a}^\dagger\hat{a})}_\infty$, respectively. The rate-equation formalism developed in the main text therefore determines the temporal evolution of the experimentally relevant first- and second-order correlation functions $g^{(1)}(\tau)$ and $g^{(2)}(\tau)$ through the quantum regression theorem.

\bibliography{refs}

@book{Petruccione_book,
    author = {Breuer, Heinz-Peter and Petruccione, Francesco},
    title = {{The Theory of Open Quantum Systems}},
    publisher = {Oxford University Press},
    year = {2007},
    month = {01},
    isbn = {9780199213900},
    doi = {10.1093/acprof:oso/9780199213900.001.0001},
    url = {https://doi.org/10.1093/acprof:oso/9780199213900.001.0001},
}

@book{Reichl_book16,
    author = {Reichl, Linda E},
    title = {{A Modern Course in Statistical Physics}},
    publisher = {Whiley-VCH},
    year = {2016},
    month = {04},
    isbn = {978-3-527-41349-2},
    url = {https://www.wiley-vch.de/de/fachgebiete/naturwissenschaften/a-modern-course-in-statistical-physics-978-3-527-41349-2},
}

@article{EP-review,
author = {El-Ganainy, Ramy and Makris, Konstantinos G and Khajavikhan, Mercedeh and Musslimani, Ziad H and Rotter, Stefan and Christodoulides, Demetrios N},
title = {{Non-Hermitian physics and PT symmetry}},
journal = {Nature Phys.},
volume = {14},
pages = {11-19},
year = {2018},
doi = {https://doi.org/10.1038/nphys4323}
}

@article{EP-review_2,
author = {Mohammad-Ali Miri  and Andrea Alù},
title = {{Exceptional points in optics and photonics}},
journal = {Science},
volume = {363},
number = {6422},
pages = {eaar7709},
year = {2019},
doi = {10.1126/science.aar7709},
URL = {https://www.science.org/doi/abs/10.1126/science.aar7709}
}

@misc{Redmann26,
    author = {Andreas Redmann and Franck Vewinger and Martin Weitz},
    title = {},
    note = {Private communications, unpublished},
    year = {2026}
}

@article{Wu19,
  title={{Observation of parity-time symmetry breaking in a single-spin system}},
  author={Wu, Y. and Liu, W. and Geng, J. and Song, X. and Ye, X. and Duan, C.-K. and Rong, X. and Du, J.},
  journal={Science},
  volume={364},
  pages={878},
  year={2019},
  doi={https://doi.org/10.1126/science.aaw8205},
}

@article{Wang24,
  title = {Exceptional Nexus in Bose-Einstein Condensates with Collective Dissipation},
  author = {Wang, Chenhao and Li, Nan and Xie, Jin and Ding, Cong and Ji, Zhonghua and Xiao, Liantuan and Jia, Suotang and Yan, Bo and Hu, Ying and Zhao, Yanting},
  journal = {Phys. Rev. Lett.},
  volume = {132},
  issue = {25},
  pages = {253401},
  numpages = {6},
  year = {2024},
  month = {Jun},
  publisher = {American Physical Society},
  doi = {10.1103/PhysRevLett.132.253401},
  url = {https://link.aps.org/doi/10.1103/PhysRevLett.132.253401}
}

@book{Kato13,
  title={{Perturbation theory for linear operators}},
  author={Kato, Tosio},
  volume={132},
  year={2013},
  publisher={Springer Science \& Business Media},
  doi={https://doi.org/10.1007/978-3-642-66282-9}
}

@article{Turaev24,
author = {Jingwen Li  and Michael Turaev  and Masakazu Matsubara  and Kristin Kliemt  and Cornelius Krellner  and Shovon Pal  and Manfred Fiebig  and Johann Kroha },
title = {Discovery of an intrinsic non-Hermitian phase transition in a bulk condensed-matter system},
journal = {Science},
volume = {393},
number = {6816},
pages = {1152-1155},
year = {2026},
doi = {10.1126/science.ady4670}
}

@article{Weitz10,
  author = {J. Klaers and J. Schmitt and F. Vewinger and M. Weitz},
  title = {{Bose–Einstein condensation of photons in an optical microcavity}},
  journal = {Nature},
  volume = {468},
  pages = {545},
  year = {2010},
  doi = {10.1038/nature09567},
  url = {https://doi.org/10.1038/nature09567}
}

@article{Pieczarka24,
  title = {{Bose--Einstein condensation of photons in a vertical-cavity surface-emitting laser}},
  author = {Pieczarka, Maciej and G{\k{e}}bski, Marcin and Piasecka, Aleksandra N and Lott, James A and Pelster, Axel and Wasiak, Micha{\l} and Czyszanowski, Tomasz},
  journal = {Nature Photonics},
  volume = {18},
  number = {10},
  pages = {1090--1096},
  year = {2024},
  publisher = {Nature Publishing Group UK London},
  doi = {10.1038/s41566-024-01478-z},
  url = {https://doi.org/10.1038/s41566-024-01478-z}
}

@article{Ciuti13,
  title = {{Quantum fluids of light}},
  author = {Carusotto, Iacopo and Ciuti, Cristiano},
  journal = {Rev. Mod. Phys.},
  volume = {85},
  issue = {1},
  pages = {299--366},
  numpages = {0},
  year = {2013},
  month = {Feb},
  publisher = {American Physical Society},
  doi = {10.1103/RevModPhys.85.299},
  url = {https://link.aps.org/doi/10.1103/RevModPhys.85.299}
}

@article{Wouter22,
  author = {J. Bloch and I. Carusotto and M. Wouters},
  title = {{Non-equilibrium Bose–Einstein condensation in photonic systems}},
  journal = {Nat Rev Phys},
  volume = {4},
  pages = {470},
  year = {2022},
  doi = {10.1038/s42254-022-00464-0},
  url = {https://doi.org/10.1038/s42254-022-00464-0}
}

@article{Kroha21,
author = {Fahri Emre Öztürk  and Tim Lappe  and Göran Hellmann  and Julian Schmitt  and Jan Klaers  and Frank Vewinger  and Johann Kroha  and Martin Weitz },
title = {{Observation of a non-Hermitian phase transition in an optical quantum gas}},
journal = {Science},
volume = {372},
number = {6537},
pages = {88-91},
year = {2021},
doi = {10.1126/science.abe9869},
URL = {https://www.science.org/doi/abs/10.1126/science.abe9869}
}

@article{TimBode19,
  title = {{Fluctuation dynamics of an open photon Bose-Einstein condensate}},
  author = {Öztürk, Fahri Emre and Lappe, Tim and Hellmann, G\"oran and Schmitt, Julian and Klaers, Jan and Vewinger, Frank and Kroha, Johann and Weitz, Martin},
  journal = {Phys. Rev. A},
  volume = {100},
  issue = {4},
  pages = {043803},
  numpages = {8},
  year = {2019},
  month = {Oct},
  publisher = {American Physical Society},
  doi = {10.1103/PhysRevA.100.043803},
  url = {https://link.aps.org/doi/10.1103/PhysRevA.100.043803}
}

@article{Schmitt14,
  title = {{Observation of Grand-Canonical Number Statistics in a Photon Bose-Einstein Condensate}},
  author = {Schmitt, Julian and Damm, Tobias and Dung, David and Vewinger, Frank and Klaers, Jan and Weitz, Martin},
  journal = {Phys. Rev. Lett.},
  volume = {112},
  issue = {3},
  pages = {030401},
  numpages = {5},
  year = {2014},
  month = {Jan},
  publisher = {American Physical Society},
  doi = {10.1103/PhysRevLett.112.030401},
  url = {https://link.aps.org/doi/10.1103/PhysRevLett.112.030401}
}

@article{Schmitt16,
  title = {{Spontaneous Symmetry Breaking and Phase Coherence of a Photon Bose-Einstein Condensate Coupled to a Reservoir}},
  author = {Schmitt, Julian and Damm, Tobias and Dung, David and Wahl, Christian and Vewinger, Frank and Klaers, Jan and Weitz, Martin},
  journal = {Phys. Rev. Lett.},
  volume = {116},
  issue = {3},
  pages = {033604},
  numpages = {5},
  year = {2016},
  month = {Jan},
  publisher = {American Physical Society},
  doi = {10.1103/PhysRevLett.116.033604},
  url = {https://link.aps.org/doi/10.1103/PhysRevLett.116.033604} 
}

@article{Walker18,
   title={{Driven-dissipative non-equilibrium Bose–Einstein condensation of less than ten photons}},
   volume={14},
   ISSN={1745-2481},
   url={http://dx.doi.org/10.1038/s41567-018-0270-1},
   DOI={10.1038/s41567-018-0270-1},
   number={12},
   journal={Nature Physics},
   publisher={Springer Science and Business Media LLC},
   author={Walker, Benjamin T. and Flatten, Lucas C. and Hesten, Henry J. and Mintert, Florian and Hunger, David and Trichet, Aurélien A. P. and Smith, Jason M. and Nyman, Robert A.},
   year={2018},
   month=sep, pages={1173–1177} 
}

@article{Nyman24a,
  title = {{Bose–Einstein condensation of light in a semiconductor quantum well microcavity}},
  author = {Schofield, Ross C. and Fu, Ming and Clarke, Edmund and Farrer, Ian and Trapalis, Aristotelis and Dhar, Himadri S. and Mukherjee, Rick and Millard, Toby Severs and Heffernan, Jon and Mintert, Florian and Nyman, Robert A. and  Oulton, Rupert F.},
  journal = {Nature Photonics},
  volume = {18},
  pages = {1083-1089},
  year = {2024},
  doi = {https://doi.org/10.1038/s41566-024-01491-2}
}

@article{Nyman24b,
  title = {{Breakdown of Temporal Coherence in Photon Condensates}},
  author = {Tang, Yijun and Dhar, Himadri S. and Oulton, Rupert F. and Nyman, Robert A. and Mintert, Florian},
  journal = {Phys. Rev. Lett.},
  volume = {132},
  issue = {17},
  pages = {173601},
  numpages = {6},
  year = {2024},
  month = {Apr},
  publisher = {American Physical Society},
  doi = {10.1103/PhysRevLett.132.173601},
  url = {https://link.aps.org/doi/10.1103/PhysRevLett.132.173601}
}

@article{Ray25,
  title = {Stabilizing Open Photon Condensates by Ghost-Attractor Dynamics},
  author = {Abouelela, Aya and Turaev, Michael and Kramer, Roman and Janning, Moritz and Kajan, Michael and Ray, Sayak and Kroha, Johann},
  journal = {Phys. Rev. Lett.},
  volume = {135},
  issue = {5},
  pages = {053402},
  numpages = {8},
  year = {2025},
  month = {Jul},
  publisher = {American Physical Society},
  doi = {10.1103/hcsq-dwcg},
  url = {https://link.aps.org/doi/10.1103/hcsq-dwcg}
}

@article{Koch24,
  title = {{Ghost Channels and Ghost Cycles Guiding Long Transients in Dynamical Systems}},
  author = {Koch, D. and Nandan, A. and Ramesan, G. and Tyukin, I. and Gorban, A. and Koseska, A.},
  journal = {Phys. Rev. Lett.},
  volume = {133},
  issue = {4},
  pages = {047202},
  numpages = {6},
  year = {2024},
  month = {Jul},
  publisher = {American Physical Society},
  doi = {10.1103/PhysRevLett.133.047202},
  url = {https://link.aps.org/doi/10.1103/PhysRevLett.133.047202} 
}

@book{Strogatz-book,
  title={{Nonlinear dynamics and chaos: with applications to physics, biology, chemistry, and engineering (studies in nonlinearity)}},
  author={Strogatz, Steven H},
  volume={1},
  year={2001},
  publisher={Westview Press}
}

@article{Strogatz89,
  title = {{Predicted power laws for delayed switching of charge-density waves}},
  author = {Strogatz, Steven H. and Westervelt, Robert M.},
  journal = {Phys. Rev. B},
  volume = {40},
  issue = {15},
  pages = {10501--10508},
  numpages = {0},
  year = {1989},
  month = {Nov},
  publisher = {American Physical Society},
  doi = {10.1103/PhysRevB.40.10501},
  url = {https://link.aps.org/doi/10.1103/PhysRevB.40.10501}
}

@article{Keeling13,
  title = {{Nonequilibrium Model of Photon Condensation}},
  author = {Kirton, Peter and Keeling, Jonathan},
  journal = {Phys. Rev. Lett.},
  volume = {111},
  issue = {10},
  pages = {100404},
  numpages = {5},
  year = {2013},
  month = {Sep},
  publisher = {American Physical Society},
  doi = {10.1103/PhysRevLett.111.100404},
  url = {https://link.aps.org/doi/10.1103/PhysRevLett.111.100404}
}

@article{Keeling15,
  title = {{Thermalization and breakdown of thermalization in photon condensates}},
  author = {Kirton, Peter and Keeling, Jonathan},
  journal = {Phys. Rev. A},
  volume = {91},
  issue = {3},
  pages = {033826},
  numpages = {15},
  year = {2015},
  month = {Mar},
  publisher = {American Physical Society},
  doi = {10.1103/PhysRevA.91.033826},
  url = {https://link.aps.org/doi/10.1103/PhysRevA.91.033826}
}

@article{Keeling17,
  title = {Suppressing and Restoring the Dicke Superradiance Transition by Dephasing and Decay},
  author = {Kirton, Peter and Keeling, Jonathan},
  journal = {Phys. Rev. Lett.},
  volume = {118},
  issue = {12},
  pages = {123602},
  year = {2017},
  publisher = {American Physical Society},
  doi = {10.1103/PhysRevLett.118.123602},
  url = {https://link.aps.org/doi/10.1103/PhysRevLett.118.123602}
}

@article{Pelster18,
doi = {10.1088/1367-2630/aac2a6},
url = {https://dx.doi.org/10.1088/1367-2630/aac2a6},
year = {2018},
month = {may},
publisher = {IOP Publishing},
volume = {20},
number = {5},
pages = {055014},
author = {Radonjić, Milan and Kopylov, Wassilij and Balaž, Antun and Pelster, Axel},
title = {{Interplay of coherent and dissipative dynamics in condensates of light}},
journal = {New Journal of Physics}
}

@article{Schmitt23,
  author = {Fahri Emre Öztürk  and Frank Vewinger  and Martin Weitz  and Julian Schmitt},
  title = {{Fluctuation-Dissipation Relation for a Bose-Einstein Condensate of Photons}},
  journal = {Phys. Rev. Lett.},
  volume = {130},
  issue = {3},
  pages = {033602},
  numpages = {6},
  year = {2023},
  month = {Jan},
  publisher = {American Physical Society},
  doi = {10.1103/PhysRevLett.130.033602},
  url = {https://link.aps.org/doi/10.1103/PhysRevLett.130.033602}
}

@phdthesis{Turaev26,
    doi = {https://doi.org/10.48565/bonndoc-770},
    author = {Michael Turaev},
    title = {Non-Equilibrium Phenomena in Correlated Electron Systems and Photon Bose-Einstein Condensates},
    school = {Rheinische Friedrich-Wilhelms-Universität Bonn},
    year = 2026,
    month = feb,
    url = {https://hdl.handle.net/20.500.11811/13862}
}

@phdthesis{Abouelela26,
    author = {{Aya Abouelela}},
    title = {Non-equilibrium dynamics in multi-mode and multi-cavity photon resonators},
    school = {Rheinische Friedrich-Wilhelms-Universität Bonn},
    year = 2026,
    month = sep,
    url = {https://hdl.handle.net/20.500.11811/14559}
}

@article{Zenodo-data,
  author       = {Janning, Moritz and
                  Kramer, Roman and
                  Turaev, Michael and
                  Ray, Sayak and
                  Kroha, Johann},
  title        = {{Figure data for ``Emergent thermal fluctuations and non-Hermitian phase transitions in open photon condensates"}},
  year         = 2026,
  journal      = {Zenodo,},
  publisher    = {Zenodo},
  doi          = {10.5281/zenodo.19097561},
  url          = {https://doi.org/10.5281/zenodo.19097561}
}

@article{roccati2026perspective,
  title={{Perspective: Interactions and Nonlinearity in Non-Hermitian Physics}},
  author={Roccati, Federico and Balducci, Federico},
  journal={\href{https://arxiv.org/abs/2603.13526}{arXiv:2603.13526}},
  year={2026}
}

@misc{Klars2026,
      title={Ghost State of Light}, 
      author={R. M. de Boer and C. Toebes and Jan Klaers and S. R. K. Rodriguez},
      year={2026},
      eprint={2605.13584},
      archivePrefix={arXiv},
      primaryClass={physics.optics},
      url={https://arxiv.org/abs/2605.13584}, 
}

\end{document}